\documentclass[aps,pra, superscriptaddress, reprint, longbibliography]{revtex4-1}

\usepackage{graphicx}
\usepackage{soul}
\usepackage[colorlinks,
			linkcolor=blue,
			urlcolor=blue,
			citecolor=blue]{hyperref}
\usepackage[T1]{fontenc}       
\usepackage[version=3]{mhchem}
\usepackage{titlesec}
\titlespacing*{\section}{0pt}{1.1\baselineskip}{\baselineskip}

\begin{document}
		
	\title[]{Programmable Mechanical Resonances in MEMS\\by Localized Joule Heating of Phase Change Materials}
	
	\author{Nicola \surname{Manca}}
	\affiliation{\mbox{Physics Department, University of Genova, Via Dodecaneso 33, Genova, 16146, Italy}}
	\affiliation{\mbox{CNR-SPIN, Corso Perrone 24 Genova, 16152, Italy}}
	
	\author{Luca \surname{Pellegrino}}
	\email{luca.pellegrino@spin.cnr.it}
	\affiliation{\mbox{CNR-SPIN, Corso Perrone 24 Genova, 16152, Italy}}
	
	\author{Teruo \surname{Kanki}}
	\affiliation{\mbox{Institute of Scientific and Industrial Research, Osaka University, Ibaraki, Osaka 567-0047, Japan}}
	
	\author{Syouta \surname{Yamasaki}}
	\affiliation{\mbox{Institute of Scientific and Industrial Research, Osaka University, Ibaraki, Osaka 567-0047, Japan}}
	
	\author{Hidekazu \surname{Tanaka}}
	\affiliation{\mbox{Institute of Scientific and Industrial Research, Osaka University, Ibaraki, Osaka 567-0047, Japan}}
	
	\author{Antonio Sergio \surname{Siri}}
	\affiliation{\mbox{Physics Department, University of Genova, Via Dodecaneso 33, Genova, 16146, Italy}}
	\affiliation{\mbox{CNR-SPIN, Corso Perrone 24 Genova, 16152, Italy}}
	
	\author{Daniele \surname{Marr\'e}}
	\affiliation{\mbox{Physics Department, University of Genova, Via Dodecaneso 33, Genova, 16146, Italy}}
	\affiliation{\mbox{CNR-SPIN, Corso Perrone 24 Genova, 16152, Italy}}	
	
	\maketitle
	\noindent This is the peer reviewed version of the following article: ``N. Manca et al. Adv. Mater. 2013, 25, 6430-6435'', which has been published in final form at \url{http://dx.doi.org/10.1002/adma.201302087}. This article may be used for non-commercial purposes in accordance with Wiley Terms and Conditions for Self-Archiving. Published online 27 August 2013\\
	
	Dynamic control of resonance frequencies of mechanical resonators is an important and challenging task for emerging applications of micro/nanoelectromechanical systems (MEMS/NEMS). The eigenfrequency of a micro/nanocantilever depends on the physical properties of its constituent materials, on mass adsorption and on interactions with external fields or internal stresses \cite{Eom2011, Waggoner2007, Li2007a}. Electrostatic forces \cite{Unterreithmeier2009} and thermal expansion in clamped geometries through electrothermal heating \cite{Jun2006}	have been employed for dynamic frequency tuning. However, material-based approaches recently introduced a new paradigm for the development of novel adaptive devices with memory effects \cite{Kniknie2010, Merced2012}. Phase transitions (PTs) and, in particular, Metal-Insulator Transitions (MIT) in correlated oxides are characterized by coexisting electronic phases at nanoscale. In this framework, dynamics and spatial evolution of PT in Vanadium Dioxide (\ce{VO2}), a well-known oxide material whose electronic behavior at phase transition is still debated \cite{Driscoll2012, Eyert2011, Zimmers2013}, have been extensively investigated in single crystals \cite{Jones2010}, nanorods\cite{Wei2009} and thin films\cite{Takami2012a, Qazilbash2011, Kim2010} through the combined detection of electrical, optical and crystallographic properties. PT of \ce{VO2} is primarily manifested by the four orders of magnitude hysteretic resistivity drop at around $\mathrm{65\,^{\circ}C}$, however it involves almost all \ce{VO2} physical properties, such as specific heat \cite{Oh2010}, optical constants \cite{Qazilbash2009} and crystallographic properties\cite{Leroux1998}. Several studies demonstrated that PT characteristics, such as shape, width and temperature, can be tailored by stress \cite{Cao2009e}, chemical doping \cite{Takami2012} or growth conditions \cite{Brassard2005}. In particular, stress-temperature phase diagram reports the formation of different crystallographic phases \cite{Cao2010}. Recent studies on \ce{VO2}-coated cantilevers exploited its phase-coexistence to gradually tune static bending \cite{Coy2010, Cao2010b} and mechanical resonances \cite{Merced2012, Rua2012, Rua2010}. However, these approaches have several drawbacks as they involve the use of focused laser heating with difficult downscaling and have no reset capabilities, unless cooling the whole sample. We recently developed a multi-resistive states memory using \ce{VO2} microbridges \cite{Pellegrino2012} that can be read, written and erased using only current pulses, locally driving the ratio between metallic and insulating nano-domains within the	hysteretic region.
	
	In this work, we report the possibility of programming multiple eigenfrequency states of a \ce{VO2}-based microresonator using a localized source of energy as that provided by Joule effect. Localized heating gradually induces the phase transition of \ce{VO2} domains with the consequent change of Young's modulus and local stress along part of the structure. The local control of the metallic filling factor can be used to reversibly tune the mechanical resonance of \ce{VO2}-based electromechanical oscillators, where the programming current and the excitation force are applied locally, allowing an easy on-chip integration	and additional volatile memory features. This study opens perspectives for developing programmable NEMS device arrays and mechanically configurable nanostructures.
	
	\begin{figure*}
		\includegraphics[width=1.0\linewidth]{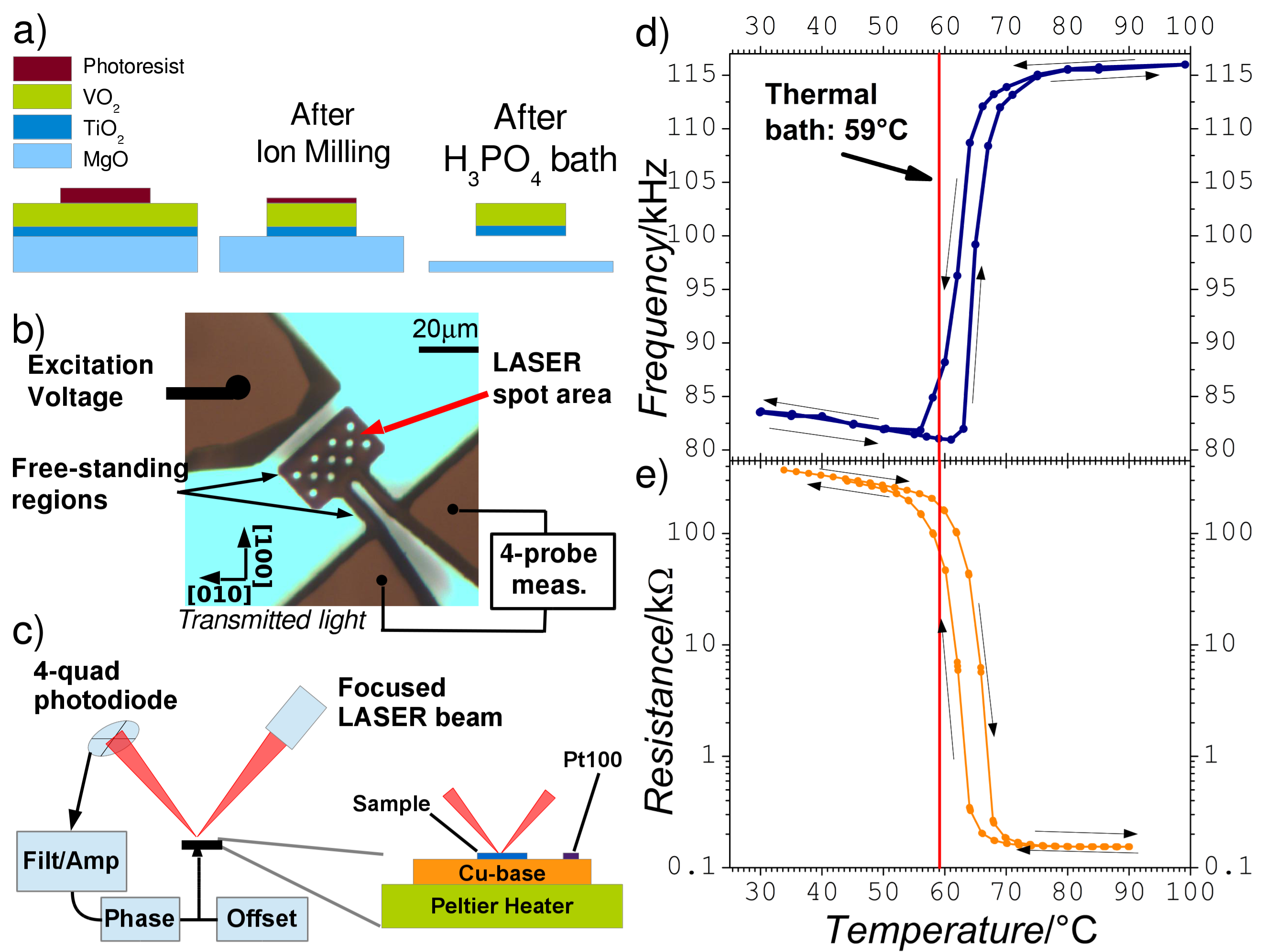}
		\caption{\label{fig:Fig1}
			Fabrication, measuring setup and characterization of cantilevers.
			a) Main steps of the fabrication process.
			b) Optical image (transmitted light) of a typical device with diagram of electrical contacts and laser spot. Darker areas are free-standing. Underetching due to acid bath is also observed on the edges of the whole pattern.
			c) Scheme of the measuring setup.
			d) Frequency vs temperature and e) resistance vs temperature plots. Resistance drop is approximately 3.5 orders of magnitude and is weakly smoothed by laser-induced local heating.
		}
	\end{figure*}
		
	We fabricated \ce{VO2}(175\,nm)/\ce{TiO2}(35\,nm) free-standing devices using standard optical lithography technique (Figure\,\ref{fig:Fig1}a). The \ce{TiO2(110)} thin film buffer layer is necessary to achieve high crystalline quality \ce{VO2} films, which cannot be obtained by depositing directly on the \ce{MgO(001)} substrate. The use of a full-oxide-based device allows integration with other oxide materials having different physical properties and the realization of high crystallographic quality ultra-thin film heterostructures \cite{Schlom2008}. \ce{TiO2(110)} films grow with the [001] direction oriented along [110] and [1-10] directions of \ce{MgO(001)}, creating a pattern of orthogonal nanodomains \cite{Okimura2005}; \ce{VO2} films grown on \ce{TiO2(110)}	show only (ll0) peaks \cite{Pellegrino2012, Muraoka2002d} and thus two in-plane orientations of strained nanodomains are expected due to the heteroepitaxial growth of \ce{VO2} on \ce{TiO2} crystals. Nanodomains structure of the films determines smooth and broad PT, contrary to what is observed when depositing on \ce{TiO2(100)}\cite{Takami2012a} substrate. We also note that device pattern is aligned along the [110] direction of MgO substrate (rotated by 45$^{\circ}$, see Figure\,\ref{fig:Fig1}b), thus \ce{TiO2} nano-domains are aligned in parallel or orthogonally with the cantilever length.
	
	Electrical resistance of the device is measured using four-probe technique controlling the bias current flowing in the free-standing area, whose geometry maximizes both capacitive coupling with the frontal gate and laser reflecting region ($\mathrm{20\,\times\,33\,\mu m}$) (Figure\,\ref{fig:Fig1}b). The holes are fabricated to speed up the release process during the acid bath. Resonance frequencies are measured using a closed-loop feedback circuit as sketched in Figure\,\ref{fig:Fig1}c. The cantilever is set oscillating at its eigenfrequency through capacitive coupling by sending the output signal of the photodiode, properly tuned in phase and amplitude, to the frontal excitation gate. A constant voltage bias is also summed to the driving voltage to avoid frequency doubling due to square dependence of the force with respect to the electric field. Electrostatic actuation is possible due to the slight gate-pad misalignment and to the breaking of electric field symmetry given by the dielectric substrate \cite{Biasotti2013}. Measurements are made in vacuum ($\mathrm{2\cdot10^{-3}\,mbar}$).
	
	\begin{figure}
		\includegraphics[width=1.0\linewidth]{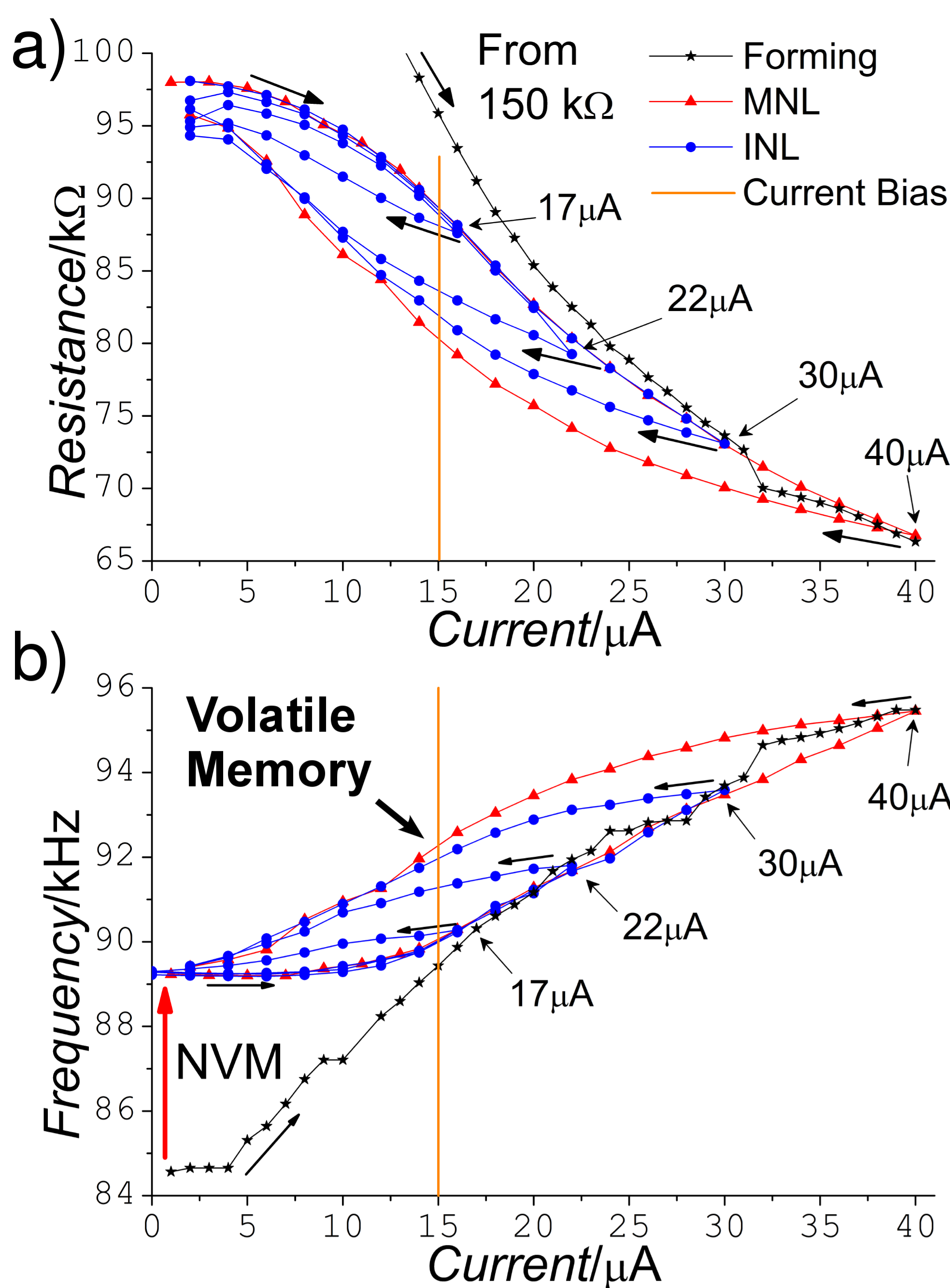}
		\caption{\label{fig:Fig2}
			Nested hysteretic loops driven by Joule self-heating.
			Resistance a) and Frequency b) vs Current loops. Plots show initial forming (black line), MNL (red line) (spanning from $\mathrm{0\,\mu A}$ to $\mathrm{40\,\mu A}$) and INL (blue lines). Inner loops are obtained limiting the current below $I_{max}$ and going back to zero. Sample temperature is set to $\mathrm{59\,^{\circ}C}$. Vertical red arrows indicate NVM effect. Volatile memory effects are instead along the vertical orange line.
		}
	\end{figure}
	
	Figures\,\ref{fig:Fig1}d and \ref{fig:Fig1}e show the temperature dependence of the device resonance frequency and electrical resistance, respectively. At PT the resistance varies by more than 3 orders of magnitude in a symmetric and reproducible curve, marking good \ce{VO2} quality. Resonance frequency varies of approximately 40\,\%, this value is very high compared to the literature \cite{Rua2012}, even considering clamped geometries \cite{Merced2012}. The red line in Figures\,\ref{fig:Fig1}d and \ref{fig:Fig1}e is the temperature of the thermal bath (copper basement in contact with the Peltier heater) where we tested the memory capabilities. Figure\,\ref{fig:Fig2}a and \ref{fig:Fig2}b show simultaneous resistance and resonance frequency measurements as a function of the current flowing into the device. Joule heating is maximum nearby the resonator arms due to the favorable aspect ratio, but	temperature profile is also determined by the balance between heat dissipation with ambient and thermal conduction along the device (see Figure\,S6). The first ``forming'' current ramp induces the largest variations of both resistance and frequency (black line). The hysteretic nature of the system emerges when decreasing the current from $I_{max}$ (in this case $\mathrm{40\,\mu A}$) to zero, where a returning path significantly different from the initial one is observed. Similarly to what happens to the electrical
	resistance \cite{Pellegrino2012}, the device mechanical eigenfrequency ($\mathrm{1^{st}}$ flexural mode), measured cycling the current from zero to $I_{max}$, defines a Main Nested Loop (MNL) (red line). Each MNL is well reproducible and its shape is determined only by the temperature bias and $I_{max}$. If the current flowing into the device overtakes $I_{max}$, a new MNL with a different shape is observed together with a lower/higher value of resistance/eigenfrequency at zero bias. The zero bias states constitute the non-volatile memory effect (NVM) and can be erased only by cooling the whole device below the hysteretic region. If the current ramp is stopped below $I_{max}$ we obtain a series of reproducible Inner Nested Loops (INL) having common onward branches, but different returning paths. Eigenfrequency and electrical resistance measured under a given current bias $I_{bias}$ (i.e. orange line in
	Figures\,\ref{fig:Fig2}a and \ref{fig:Fig2}b) can thus have multiple branch-dependent values. These multiple values constitute the volatile memory of our device, because the states can be erased just tuning the current bias, without cooling the device. The read current ($I_{bias}$) should be chosen carefully: if it is too close to zero or $I_{max}$, the available states are almost indistinguishable. The use of current bias, instead of voltage, allows fine control of the PT due to the negative differential resistance upon heating \cite{Tselev2011} avoiding catastrophic effects and improving device lifetime \cite{Crunteanu2010}.
	
	\begin{figure*}
		\includegraphics[width=1.0\linewidth]{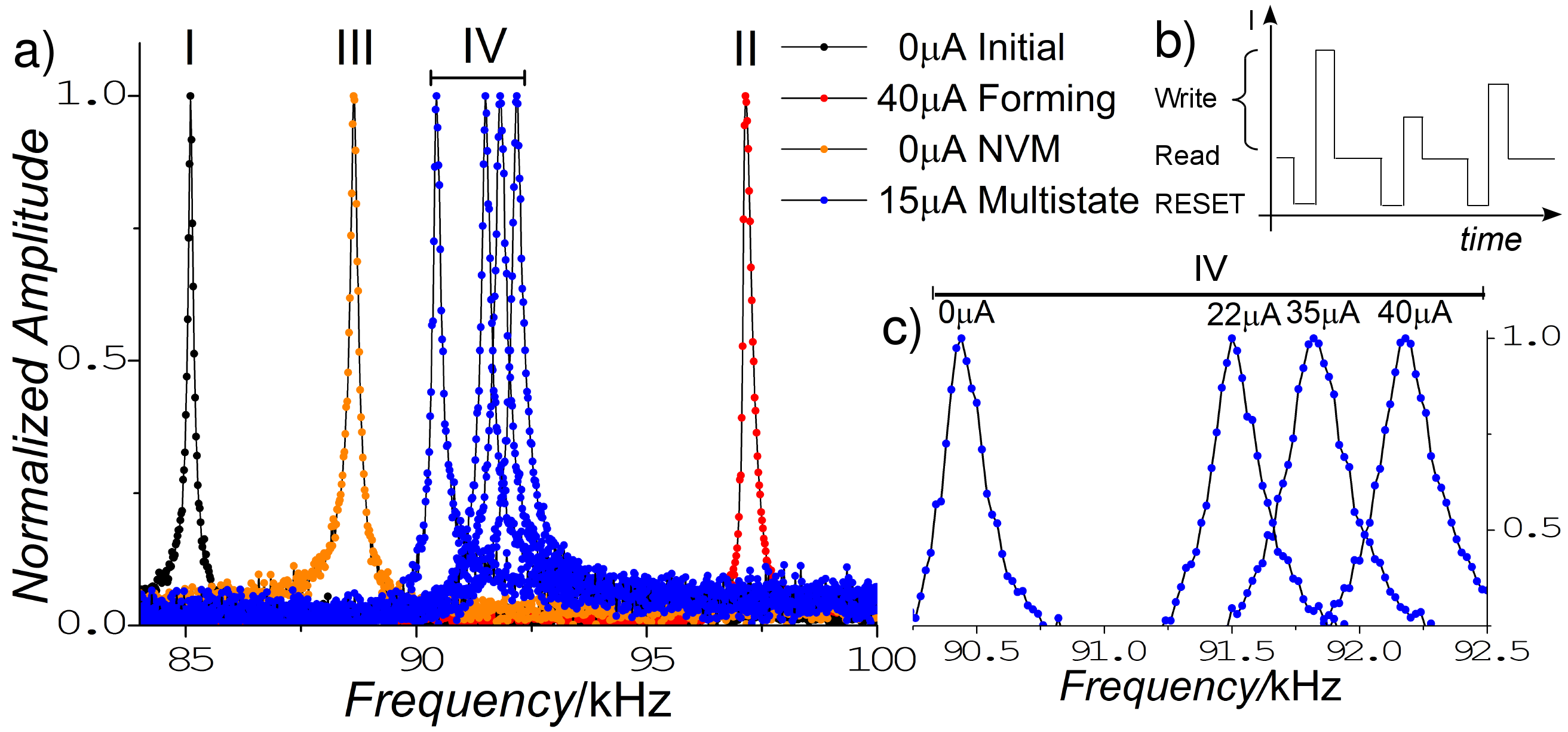}
		\caption{\label{fig:Fig3}
			Spectral analysis with different current biases.
			a) Spectral response of our device ($\mathrm{1^{\circ}}$ flexural mode) in different frequency states:
			(I) just heated to $\mathrm{59\,^{\circ}C}$ and $\mathrm{0\,\mu A}$ bias current applied.
			(II) $\mathrm{40\,\mu A}$ bias current (forming process).
			(III) $\mathrm{0\,\mu A}$ bias current after the forming process.
			(IV) $\mathrm{15\,\mu A}$ bias current states, each peak corresponds to a frequency state programmed using different write current pulses.
			b) Diagram of the write/read pulse sequence used for writing states (IV) magnified in figure c) where each peak is labelled with the corresponding writing pulse value.
		}
	\end{figure*}
	
\begin{figure*}
		\includegraphics[width=1.0\linewidth]{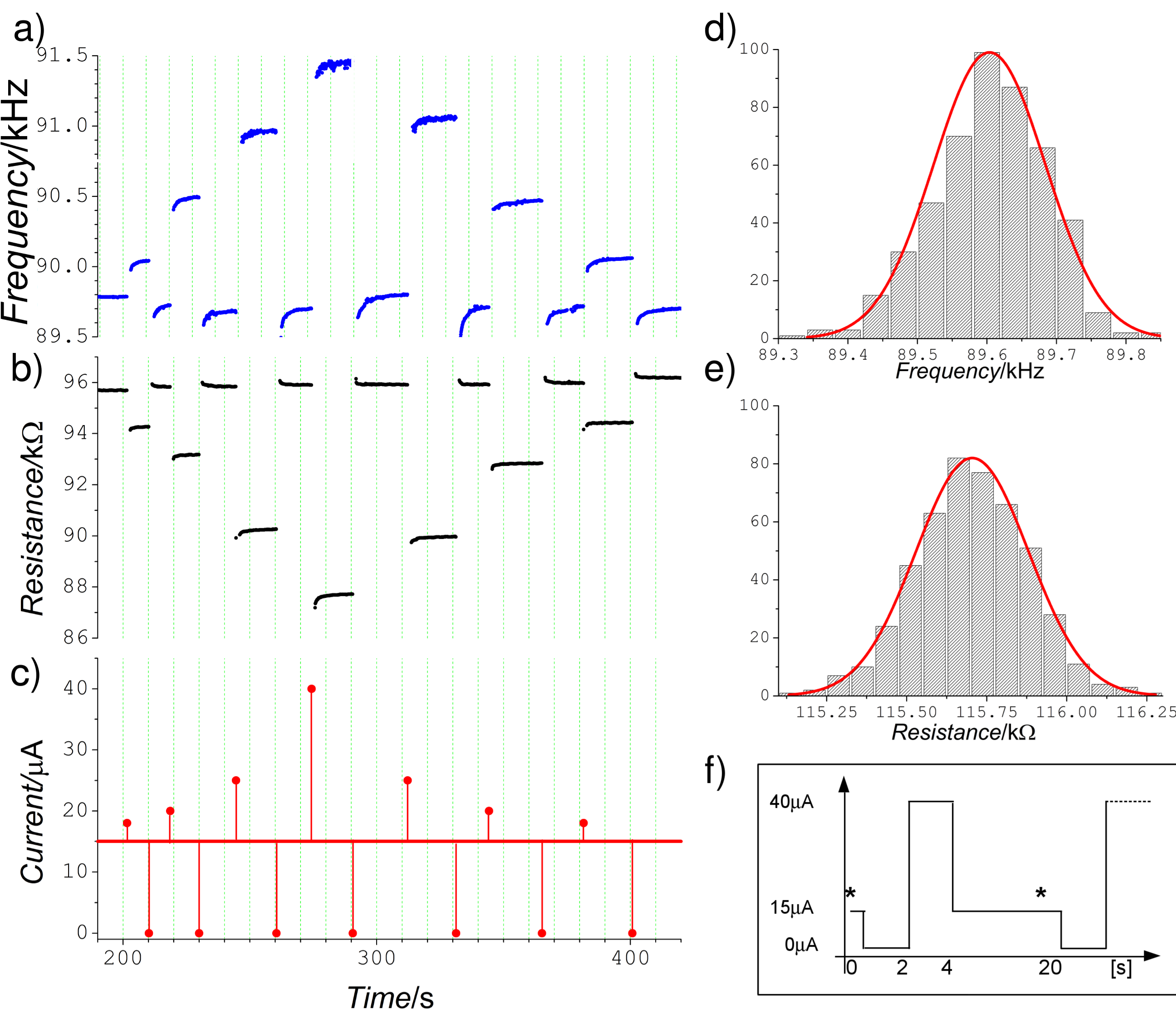}
		\caption{\label{fig:Fig4}
			Time plot and dispersion of mechanical and electrical programmed states.
			Time plot of a) mechanical eigenfrequency and b) electrical resistance states written by different current pulses c).
			Zero current pulses (RESET) restore the ``initial'' state. Reading is made with $\mathrm{15\,\mu A}$ current bias
			and $\mathrm{59\,^{\circ}C}$ thermal bath temperature.
			Histograms of d) frequency and e) resistance states upon a series of programming cycle showed in f). Y scale is counts. Bin widths are 40\,Hz and 75\,$\Omega$, respectively. Lines are Gaussian fit, asterisks in f) mark triggered measurement.
			Data in d) and e) have been collected under slightly different substrate temperature.
		}
	\end{figure*}

	Figure\,\ref{fig:Fig3}a shows normalized frequency spectra of the device in different programmed states. Each peak is numbered according to the following sequence of operations. The system is initially in the pre-forming state (I). The eigenfrequency increases during the current ramp to $\mathrm{40\,\mu A}$ bias ($I_{max}$) (II) and when the bias current is set again to zero it collapses to a different value (III) (NVM contribution). The device is then biased with $\mathrm{15\,\mu A}$ read current and multi-eigenfrequency states (blue peaks) can be selected by current pulses of amplitudes between $\mathrm{15\,\mu A}$ and $\mathrm{40\,\mu A}$ (IV) (Figure\,\ref{fig:Fig3}b–c). A zero-current pulse acts
	as RESET, restoring the initial state. Frequency vs Pulse amplitude relationship is clearly non-linear, as expected considering measurements of Figure\,\ref{fig:Fig2}b. Memory retention is tested in the time plot of Figure\,\ref{fig:Fig4}. After the initial procedure (thermal bath at $\mathrm{59\,^{\circ}C~\rightarrow~40\,\mu A}$ forming pulse $\mathrm{\rightarrow~0\,\mu A}$ pulse $\mathrm{\rightarrow~15\,\mu A}$ bias) we recorded the eigenfrequency (Figure\,\ref{fig:Fig4}a) and resistance (Figure\,\ref{fig:Fig4}b) while sending a series of current pulses (17, 20, 25 and 40\,$\mu$A), each one preceded by a RESET (Figure\,\ref{fig:Fig4}c). Our tests clearly evidence the retention properties of our devices and the possibility of obtaining distinguished multiple eigenfrequency states. The drifts measured for both frequency and resistance soon after current pulses are too slow to be due to the measurement setup and the estimated thermal time constant of our devices are several order of magnitude smaller. Similar drifts have been observed in VO2-based devices and accounted to thermal relaxation \cite{Coy2010} as well as in other phase change materials where have been related to structural relaxation at PT \cite{Ielmini2007}. We also suppose that the origin of such drifts is linked to temperature-related processes of domains evolution,	as they are observed also when heating with a laser pulse of higher power, but are absent when simply inverting the current sign thus excluding motion of charges or ions. So far, we could not observe any discrete behavior in the frequency vs current plots, this is the fingerprint of the fine domain structure of our films grown on \ce{TiO2(110)}. We interpret the small deviations of eigenfrequency and resistance around each programmed state as statistical fluctuation of the number, position or size	of metallic domains. Also, positive deviations of resistance are
	correlated with negative deviations of the eigenfrequency (see Figure\,S4 and S5, Supporting Information). Reproducibility
	of programmed states was tested using a series of write/erase cycles: ``$0\,\mu$A RESET $\mathrm{\rightarrow~40\,\mu A}$ pulse write $\mathrm{\rightarrow~15\,\mu A}$ read'' and histograms of dispersion for both resistive and frequency programmed states are reported in Figures\,\ref{fig:Fig4}d and \ref{fig:Fig4}e (see also Figure\,S4). We estimated the programming reproducibility as the FWHM of Figures\ref{fig:Fig4}d and \ref{fig:Fig4}e, giving 400\,$\Omega$ and 200\,Hz values. At the PT of a single nanodomain both resistive and structural effects occur. Structural phase transition in \ce{VO2} involves change of the lattice parameters, determining a local stress in the neighborhood of the transited domain. In bulk \ce{VO2} during monoclinic to rutile PT the c-axis is contracted by $\mathrm{\approx0.8\,\%}$ while a and b axis are elongated by $\mathrm{\approx0.5\,\%}$ \cite{Eyert2011} and 1\,GPa stress across PT in \ce{VO2}-coated Si cantilevers have been reported \cite{Rua2010}. In our microstructure in-plane stresses associated to PT partially compensate due to the (110) orientation of \ce{VO2}-rutile unit cell and the inferred orthogonal in-plane orientations of \ce{VO2} nanodomains. Resulting in-plane stress produces the observed upward bending of the device during PT and upon current bias, but only small difference in the optical images is observed when comparing high and low temperature states (see Figures\,S2 and\,S3). This indicates that phase coexistence between metallic and insulating clusters significantly contributes to bending, especially if it occurs along film thickness. The contribution from changes of intrinsic material properties and that of PT-induced internal stresses (and related deformations) to the observed frequency shifts during device operation are still under study. A simple model involving only a change of Young's modulus would require its increase in the high temperature phase of a factor of about 2.5, but this values is in contrast to what reported by several works in literature\cite{Rua2012, Fan2009}. Thus,	stress plays a relevant role, first because of stress-stiffening effects and secondly because of related structural deformations that increase rigidity of the flexural modes.
	
	In summary, we have showed how the synergy between correlated electron materials exhibiting phase transition and microelectromechanical systems may open new possibilities for developing programmable (nano)mechanical devices. Using vanadium dioxide, we demonstrated that we can program reversibly the overall mechanical properties, such as the eigenfrequency of a microresonator, over a wide range of different stable values using the integrated control provided by confined heating given by Joule effect. Multistate programming is possible by controlling the PT of \ce{VO2} nanodomains with
	progressive modification of internal stresses and downscaling is limited by the minimum size of a stable domain that can be addressed. The high sensitivities given by nanomechanical detection methods \cite{Rugar2004, OConnell2010, Naik2009a, Sengupta2010} open perspectives for using oxide-based resonators as tools for studying nanoscale phase separation phenomena in confined geometries. Others strongly correlated oxide materials, such as manganites, high-T$_{\mathrm{c}}$ superconductors or oxide-based heterostructures have also promising potential in view of a correlated oxide nanomechanics.

	\section*{Experimental Section}
	
	\indent\indent \textit{Thin film deposition and device fabrication:} Thin films were fabricated using Pulsed Laser Deposition (PLD) in oxygen atmosphere: 1.0\,Pa for \ce{VO2} and 0.1\,Pa for \ce{TiO2} at substrate temperature 460\,$^{\circ}$C and 680\,$^{\circ}$C, respectively, with 18\,mJ\,cm$^{-2}$ laser fluency and 2\,Hz repetition rate. The measured growth rate for \ce{VO2} and \ce{TiO2} was respectively 20\,nm\,h$^{−1}$	and 30\,nm\,h$^{−1}$. Device patterns were fabricated using standard microlithographic technique. \ce{VO2/TiO2} heterostructures were patterned into the desired geometries by Ion milling (500\,eV, 0.2\,mA\,cm$^{-2}$, 60\,min). Free-standing structures were obtained by soaking the sample in \ce{H3PO4} 8.5\% acid bath for 6 hours; this step progressively removes	portion of the MgO substrate starting from the edges of the pattern and leaves the \ce{TiO2} and \ce{VO2} films undamaged. The holes (3\,$\mu$m diameter) on the cantilever are fabricated to help the release process. Samples are dried from the wet etching bath using a Critical Point Drier (CPD) to avoid stiction. We fabricated cantilevers having typically 20\,$\mu$m long and 33\,$\mu$m wide pads; arms are 5\,$\mu$m of width and 15\,$\mu$m of length; distance between the arms is 3\,$\mu$m, while separation between the frontal gate pad and the cantilever is 5\,$\mu$m.
	
	\textit{Electrical and mechanical measurements:} Measurements are performed in a vacuum chamber at low vacuum pressure ($\mathrm{2\cdot10^{-3}}$\,mbar). Samples are glued with silver paste on a copper stage that is heated using the PC-controlled Peltier element. Temperature is measured by a Pt100 thermometer and its noise is less than 0.1\,K. Mechanical resonance frequency is detected by optical lever technique, using a four-quadrants photodiode. The output signal from the photodiode is filtered and amplified and then sent to an electronic circuit that allows the user to tune phase shift, DC offset and signal amplitude in order to maintain the cantilever at its eigenfrequency with constant oscillation amplitude. Resonance frequency is read by a frequency counter (Keithley 3390). Data	acquisition, temperature and electrical signals are all controlled using a
	LABView\textregistered  interface. Laser source is a ``laser micro focus generator'' from Shafter+Kirchhoff, model 13MC-M100-95CM-635 with maximum power of 10\,mW and 5\,$\mu$m of focused spot diameter. Laser spot has been focused over the whole pad area, minimizing the inward power density. Laser power can be tuned by external trimmer in the $0\rightarrow10$\,mW range and was chosen in order to avoid any relevant resistive and frequency
	shift, equivalent to less than 5\,$\mu$A current bias and can be considered as a constant additive power source. To test temperature-induced drift we used an higher intensity, equivalent to shifting resistance/frequency by 40\,$\mu$A current bias.
	
	\section*{Acknowledgements}
	We acknowledge financial support from FIRB RBAP115AYN ``Oxides at the nanoscale: multifunctionality and applications'', Fondazione CARIGE, PRIN 2010NR4MXA ``OXIDE'', Progetti di Ateneo 2011, Grants-in-Aid for Scientific Research B (No.25286058) in Japan Society for promotion of Science (JSPS), Grant-in-Aid for Young Scientists S (Grant No. 21676001) from the Ministry of Education, Culture, Sports, Science, and Technology (MEXT). Useful discussions with Michele Biasotti, Renato Buzio, Riccardo Mazzarello and technical help from Roberto Chittofrati and Manuele Gargano are also acknowledged. L. P. also thanks Takuya Matsumoto for instructive discussions about frequency modulation AFM techniques.
	
	\bibliographystyle{apsrev4-1}
	\bibliography{library.bib}	

\begin{thebibliography}{39}%
\makeatletter
\providecommand \@ifxundefined [1]{%
 \@ifx{#1\undefined}
}%
\providecommand \@ifnum [1]{%
 \ifnum #1\expandafter \@firstoftwo
 \else \expandafter \@secondoftwo
 \fi
}%
\providecommand \@ifx [1]{%
 \ifx #1\expandafter \@firstoftwo
 \else \expandafter \@secondoftwo
 \fi
}%
\providecommand \natexlab [1]{#1}%
\providecommand \enquote  [1]{``#1''}%
\providecommand \bibnamefont  [1]{#1}%
\providecommand \bibfnamefont [1]{#1}%
\providecommand \citenamefont [1]{#1}%
\providecommand \href@noop [0]{\@secondoftwo}%
\providecommand \href [0]{\begingroup \@sanitize@url \@href}%
\providecommand \@href[1]{\@@startlink{#1}\@@href}%
\providecommand \@@href[1]{\endgroup#1\@@endlink}%
\providecommand \@sanitize@url [0]{\catcode `\\12\catcode `\$12\catcode
  `\&12\catcode `\#12\catcode `\^12\catcode `\_12\catcode `\%12\relax}%
\providecommand \@@startlink[1]{}%
\providecommand \@@endlink[0]{}%
\providecommand \url  [0]{\begingroup\@sanitize@url \@url }%
\providecommand \@url [1]{\endgroup\@href {#1}{\urlprefix }}%
\providecommand \urlprefix  [0]{URL }%
\providecommand \Eprint [0]{\href }%
\providecommand \doibase [0]{http://dx.doi.org/}%
\providecommand \selectlanguage [0]{\@gobble}%
\providecommand \bibinfo  [0]{\@secondoftwo}%
\providecommand \bibfield  [0]{\@secondoftwo}%
\providecommand \translation [1]{[#1]}%
\providecommand \BibitemOpen [0]{}%
\providecommand \bibitemStop [0]{}%
\providecommand \bibitemNoStop [0]{.\EOS\space}%
\providecommand \EOS [0]{\spacefactor3000\relax}%
\providecommand \BibitemShut  [1]{\csname bibitem#1\endcsname}%
\let\auto@bib@innerbib\@empty
\bibitem [{\citenamefont {Eom}\ \emph {et~al.}(2011)\citenamefont {Eom},
  \citenamefont {Park}, \citenamefont {Yoon},\ and\ \citenamefont
  {Kwon}}]{Eom2011}%
  \BibitemOpen
  \bibfield  {author} {\bibinfo {author} {\bibfnamefont {K.}~\bibnamefont
  {Eom}}, \bibinfo {author} {\bibfnamefont {H.~S.}\ \bibnamefont {Park}},
  \bibinfo {author} {\bibfnamefont {D.~S.}\ \bibnamefont {Yoon}}, \ and\
  \bibinfo {author} {\bibfnamefont {T.}~\bibnamefont {Kwon}},\ }\href {\doibase
  10.1016/j.physrep.2011.03.002} {\bibfield  {journal} {\bibinfo  {journal}
  {Phys. Rep.}\ }\textbf {\bibinfo {volume} {503}},\ \bibinfo {pages} {115}
  (\bibinfo {year} {2011})}\BibitemShut {NoStop}%
\bibitem [{\citenamefont {Waggoner}\ and\ \citenamefont
  {Craighead}(2007)}]{Waggoner2007}%
  \BibitemOpen
  \bibfield  {author} {\bibinfo {author} {\bibfnamefont {P.~S.}\ \bibnamefont
  {Waggoner}}\ and\ \bibinfo {author} {\bibfnamefont {H.~G.}\ \bibnamefont
  {Craighead}},\ }\href {\doibase 10.1039/b707401h} {\bibfield  {journal}
  {\bibinfo  {journal} {Lab Chip}\ }\textbf {\bibinfo {volume} {7}},\ \bibinfo
  {pages} {1238} (\bibinfo {year} {2007})}\BibitemShut {NoStop}%
\bibitem [{\citenamefont {Li}\ \emph {et~al.}(2007)\citenamefont {Li},
  \citenamefont {Tang},\ and\ \citenamefont {Roukes}}]{Li2007a}%
  \BibitemOpen
  \bibfield  {author} {\bibinfo {author} {\bibfnamefont {M.}~\bibnamefont
  {Li}}, \bibinfo {author} {\bibfnamefont {H.~X.}\ \bibnamefont {Tang}}, \ and\
  \bibinfo {author} {\bibfnamefont {M.~L.}\ \bibnamefont {Roukes}},\ }\href
  {\doibase 10.1038/nnano.2006.208} {\bibfield  {journal} {\bibinfo  {journal}
  {Nat. Nanotechnol.}\ }\textbf {\bibinfo {volume} {2}},\ \bibinfo {pages}
  {114} (\bibinfo {year} {2007})}\BibitemShut {NoStop}%
\bibitem [{\citenamefont {Unterreithmeier}\ \emph {et~al.}(2009)\citenamefont
  {Unterreithmeier}, \citenamefont {Weig},\ and\ \citenamefont
  {Kotthaus}}]{Unterreithmeier2009}%
  \BibitemOpen
  \bibfield  {author} {\bibinfo {author} {\bibfnamefont {Q.~P.}\ \bibnamefont
  {Unterreithmeier}}, \bibinfo {author} {\bibfnamefont {E.~M.}\ \bibnamefont
  {Weig}}, \ and\ \bibinfo {author} {\bibfnamefont {J.~P.}\ \bibnamefont
  {Kotthaus}},\ }\href {\doibase 10.1038/nature07932} {\bibfield  {journal}
  {\bibinfo  {journal} {Nature}\ }\textbf {\bibinfo {volume} {458}},\ \bibinfo
  {pages} {1001} (\bibinfo {year} {2009})}\BibitemShut {NoStop}%
\bibitem [{\citenamefont {Jun}\ \emph {et~al.}(2006)\citenamefont {Jun},
  \citenamefont {Huang}, \citenamefont {Manolidis}, \citenamefont {Zorman},
  \citenamefont {Mehregany},\ and\ \citenamefont {Hone}}]{Jun2006}%
  \BibitemOpen
  \bibfield  {author} {\bibinfo {author} {\bibfnamefont {S.~C.}\ \bibnamefont
  {Jun}}, \bibinfo {author} {\bibfnamefont {X.~M.~H.}\ \bibnamefont {Huang}},
  \bibinfo {author} {\bibfnamefont {M.}~\bibnamefont {Manolidis}}, \bibinfo
  {author} {\bibfnamefont {C.~a.}\ \bibnamefont {Zorman}}, \bibinfo {author}
  {\bibfnamefont {M.}~\bibnamefont {Mehregany}}, \ and\ \bibinfo {author}
  {\bibfnamefont {J.}~\bibnamefont {Hone}},\ }\href {\doibase
  10.1088/0957-4484/17/5/057} {\bibfield  {journal} {\bibinfo  {journal}
  {Nanotechnology}\ }\textbf {\bibinfo {volume} {17}},\ \bibinfo {pages} {1506}
  (\bibinfo {year} {2006})}\BibitemShut {NoStop}%
\bibitem [{\citenamefont {Kniknie}\ \emph {et~al.}(2010)\citenamefont
  {Kniknie}, \citenamefont {Bellouard}, \citenamefont {Homburg}, \citenamefont
  {Nijmeijer}, \citenamefont {Wang},\ and\ \citenamefont
  {Vlassak}}]{Kniknie2010}%
  \BibitemOpen
  \bibfield  {author} {\bibinfo {author} {\bibfnamefont {T.}~\bibnamefont
  {Kniknie}}, \bibinfo {author} {\bibfnamefont {Y.}~\bibnamefont {Bellouard}},
  \bibinfo {author} {\bibfnamefont {E.}~\bibnamefont {Homburg}}, \bibinfo
  {author} {\bibfnamefont {H.}~\bibnamefont {Nijmeijer}}, \bibinfo {author}
  {\bibfnamefont {X.}~\bibnamefont {Wang}}, \ and\ \bibinfo {author}
  {\bibfnamefont {J.~J.}\ \bibnamefont {Vlassak}},\ }\href {\doibase
  10.1088/0960-1317/20/1/015039} {\bibfield  {journal} {\bibinfo  {journal} {J.
  Micromechanics Microengineering}\ }\textbf {\bibinfo {volume} {20}},\
  \bibinfo {pages} {015039} (\bibinfo {year} {2010})}\BibitemShut {NoStop}%
\bibitem [{\citenamefont {Merced}\ \emph {et~al.}(2012)\citenamefont {Merced},
  \citenamefont {Cabrera}, \citenamefont {D{\'{a}}vila}, \citenamefont
  {Fern{\`{a}}ndez},\ and\ \citenamefont {Sep{\'{u}}lveda}}]{Merced2012}%
  \BibitemOpen
  \bibfield  {author} {\bibinfo {author} {\bibfnamefont {E.}~\bibnamefont
  {Merced}}, \bibinfo {author} {\bibfnamefont {R.}~\bibnamefont {Cabrera}},
  \bibinfo {author} {\bibfnamefont {N.}~\bibnamefont {D{\'{a}}vila}}, \bibinfo
  {author} {\bibfnamefont {F.}~\bibnamefont {Fern{\`{a}}ndez}}, \ and\ \bibinfo
  {author} {\bibfnamefont {N.}~\bibnamefont {Sep{\'{u}}lveda}},\ }\href
  {\doibase 10.1088/0964-1726/21/3/035007} {\bibfield  {journal} {\bibinfo
  {journal} {Smart Mater. Struct.}\ }\textbf {\bibinfo {volume} {21}},\
  \bibinfo {pages} {035007} (\bibinfo {year} {2012})}\BibitemShut {NoStop}%
\bibitem [{\citenamefont {Driscoll}\ \emph {et~al.}(2012)\citenamefont
  {Driscoll}, \citenamefont {Quinn}, \citenamefont {{Di Ventra}}, \citenamefont
  {Basov}, \citenamefont {Seo}, \citenamefont {Lee}, \citenamefont {Kim},\ and\
  \citenamefont {Smith}}]{Driscoll2012}%
  \BibitemOpen
  \bibfield  {author} {\bibinfo {author} {\bibfnamefont {T.}~\bibnamefont
  {Driscoll}}, \bibinfo {author} {\bibfnamefont {J.}~\bibnamefont {Quinn}},
  \bibinfo {author} {\bibfnamefont {M.}~\bibnamefont {{Di Ventra}}}, \bibinfo
  {author} {\bibfnamefont {D.~N.}\ \bibnamefont {Basov}}, \bibinfo {author}
  {\bibfnamefont {G.}~\bibnamefont {Seo}}, \bibinfo {author} {\bibfnamefont
  {Y.-w.}\ \bibnamefont {Lee}}, \bibinfo {author} {\bibfnamefont {H.-t.}\
  \bibnamefont {Kim}}, \ and\ \bibinfo {author} {\bibfnamefont {D.~R.}\
  \bibnamefont {Smith}},\ }\href {\doibase 10.1103/PhysRevB.86.094203}
  {\bibfield  {journal} {\bibinfo  {journal} {Phys. Rev. B}\ }\textbf {\bibinfo
  {volume} {86}},\ \bibinfo {pages} {094203} (\bibinfo {year}
  {2012})}\BibitemShut {NoStop}%
\bibitem [{\citenamefont {Eyert}(2011)}]{Eyert2011}%
  \BibitemOpen
  \bibfield  {author} {\bibinfo {author} {\bibfnamefont {V.}~\bibnamefont
  {Eyert}},\ }\href {\doibase 10.1103/PhysRevLett.107.016401} {\bibfield
  {journal} {\bibinfo  {journal} {Phys. Rev. Lett.}\ }\textbf {\bibinfo
  {volume} {107}},\ \bibinfo {pages} {016401} (\bibinfo {year}
  {2011})}\BibitemShut {NoStop}%
\bibitem [{\citenamefont {Zimmers}\ \emph {et~al.}(2013)\citenamefont
  {Zimmers}, \citenamefont {Aigouy}, \citenamefont {Mortier}, \citenamefont
  {Sharoni}, \citenamefont {Wang}, \citenamefont {West}, \citenamefont
  {Ram{\'{i}}rez},\ and\ \citenamefont {Schuller}}]{Zimmers2013}%
  \BibitemOpen
  \bibfield  {author} {\bibinfo {author} {\bibfnamefont {A.}~\bibnamefont
  {Zimmers}}, \bibinfo {author} {\bibfnamefont {L.}~\bibnamefont {Aigouy}},
  \bibinfo {author} {\bibfnamefont {M.}~\bibnamefont {Mortier}}, \bibinfo
  {author} {\bibfnamefont {A.}~\bibnamefont {Sharoni}}, \bibinfo {author}
  {\bibfnamefont {S.}~\bibnamefont {Wang}}, \bibinfo {author} {\bibfnamefont
  {K.~G.}\ \bibnamefont {West}}, \bibinfo {author} {\bibfnamefont {J.~G.}\
  \bibnamefont {Ram{\'{i}}rez}}, \ and\ \bibinfo {author} {\bibfnamefont
  {I.~K.}\ \bibnamefont {Schuller}},\ }\href {\doibase
  10.1103/PhysRevLett.110.056601} {\bibfield  {journal} {\bibinfo  {journal}
  {Phys. Rev. Lett.}\ }\textbf {\bibinfo {volume} {110}},\ \bibinfo {pages}
  {056601} (\bibinfo {year} {2013})}\BibitemShut {NoStop}%
\bibitem [{\citenamefont {Jones}\ \emph {et~al.}(2010)\citenamefont {Jones},
  \citenamefont {Berweger}, \citenamefont {Wei}, \citenamefont {Cobden},\ and\
  \citenamefont {Raschke}}]{Jones2010}%
  \BibitemOpen
  \bibfield  {author} {\bibinfo {author} {\bibfnamefont {A.~C.}\ \bibnamefont
  {Jones}}, \bibinfo {author} {\bibfnamefont {S.}~\bibnamefont {Berweger}},
  \bibinfo {author} {\bibfnamefont {J.}~\bibnamefont {Wei}}, \bibinfo {author}
  {\bibfnamefont {D.}~\bibnamefont {Cobden}}, \ and\ \bibinfo {author}
  {\bibfnamefont {M.~B.}\ \bibnamefont {Raschke}},\ }\href {\doibase
  10.1021/nl903765h} {\bibfield  {journal} {\bibinfo  {journal} {Nano Lett.}\
  }\textbf {\bibinfo {volume} {10}},\ \bibinfo {pages} {1574} (\bibinfo {year}
  {2010})}\BibitemShut {NoStop}%
\bibitem [{\citenamefont {Wei}\ \emph {et~al.}(2009)\citenamefont {Wei},
  \citenamefont {Wang}, \citenamefont {Chen},\ and\ \citenamefont
  {Cobden}}]{Wei2009}%
  \BibitemOpen
  \bibfield  {author} {\bibinfo {author} {\bibfnamefont {J.}~\bibnamefont
  {Wei}}, \bibinfo {author} {\bibfnamefont {Z.}~\bibnamefont {Wang}}, \bibinfo
  {author} {\bibfnamefont {W.}~\bibnamefont {Chen}}, \ and\ \bibinfo {author}
  {\bibfnamefont {D.~H.}\ \bibnamefont {Cobden}},\ }\href {\doibase
  10.1038/nnano.2009.141} {\bibfield  {journal} {\bibinfo  {journal} {Nat.
  Nanotechnol.}\ }\textbf {\bibinfo {volume} {4}},\ \bibinfo {pages} {420}
  (\bibinfo {year} {2009})}\BibitemShut {NoStop}%
\bibitem [{\citenamefont {Takami}\ \emph
  {et~al.}(2012{\natexlab{a}})\citenamefont {Takami}, \citenamefont {Kawatani},
  \citenamefont {Ueda}, \citenamefont {Fujiwara}, \citenamefont {Kanki},\ and\
  \citenamefont {Tanaka}}]{Takami2012a}%
  \BibitemOpen
  \bibfield  {author} {\bibinfo {author} {\bibfnamefont {H.}~\bibnamefont
  {Takami}}, \bibinfo {author} {\bibfnamefont {K.}~\bibnamefont {Kawatani}},
  \bibinfo {author} {\bibfnamefont {H.}~\bibnamefont {Ueda}}, \bibinfo {author}
  {\bibfnamefont {K.}~\bibnamefont {Fujiwara}}, \bibinfo {author}
  {\bibfnamefont {T.}~\bibnamefont {Kanki}}, \ and\ \bibinfo {author}
  {\bibfnamefont {H.}~\bibnamefont {Tanaka}},\ }\href {\doibase
  10.1063/1.4773371} {\bibfield  {journal} {\bibinfo  {journal} {Appl. Phys.
  Lett.}\ }\textbf {\bibinfo {volume} {101}},\ \bibinfo {pages} {263111}
  (\bibinfo {year} {2012}{\natexlab{a}})}\BibitemShut {NoStop}%
\bibitem [{\citenamefont {Qazilbash}\ \emph {et~al.}(2011)\citenamefont
  {Qazilbash}, \citenamefont {Tripathi}, \citenamefont {Schafgans},
  \citenamefont {Kim}, \citenamefont {Kim}, \citenamefont {Cai}, \citenamefont
  {Holt}, \citenamefont {Maser}, \citenamefont {Keilmann}, \citenamefont
  {Shpyrko},\ and\ \citenamefont {Basov}}]{Qazilbash2011}%
  \BibitemOpen
  \bibfield  {author} {\bibinfo {author} {\bibfnamefont {M.~M.}\ \bibnamefont
  {Qazilbash}}, \bibinfo {author} {\bibfnamefont {A.}~\bibnamefont {Tripathi}},
  \bibinfo {author} {\bibfnamefont {a.~a.}\ \bibnamefont {Schafgans}}, \bibinfo
  {author} {\bibfnamefont {B.-J.}\ \bibnamefont {Kim}}, \bibinfo {author}
  {\bibfnamefont {H.-T.}\ \bibnamefont {Kim}}, \bibinfo {author} {\bibfnamefont
  {Z.}~\bibnamefont {Cai}}, \bibinfo {author} {\bibfnamefont {M.~V.}\
  \bibnamefont {Holt}}, \bibinfo {author} {\bibfnamefont {J.~M.}\ \bibnamefont
  {Maser}}, \bibinfo {author} {\bibfnamefont {F.}~\bibnamefont {Keilmann}},
  \bibinfo {author} {\bibfnamefont {O.~G.}\ \bibnamefont {Shpyrko}}, \ and\
  \bibinfo {author} {\bibfnamefont {D.~N.}\ \bibnamefont {Basov}},\ }\href
  {\doibase 10.1103/PhysRevB.83.165108} {\bibfield  {journal} {\bibinfo
  {journal} {Phys. Rev. B}\ }\textbf {\bibinfo {volume} {83}},\ \bibinfo
  {pages} {165108} (\bibinfo {year} {2011})}\BibitemShut {NoStop}%
\bibitem [{\citenamefont {Kim}\ \emph {et~al.}(2010)\citenamefont {Kim},
  \citenamefont {Ko}, \citenamefont {Frenzel}, \citenamefont {Ramanathan},\
  and\ \citenamefont {Hoffman}}]{Kim2010}%
  \BibitemOpen
  \bibfield  {author} {\bibinfo {author} {\bibfnamefont {J.}~\bibnamefont
  {Kim}}, \bibinfo {author} {\bibfnamefont {C.}~\bibnamefont {Ko}}, \bibinfo
  {author} {\bibfnamefont {A.}~\bibnamefont {Frenzel}}, \bibinfo {author}
  {\bibfnamefont {S.}~\bibnamefont {Ramanathan}}, \ and\ \bibinfo {author}
  {\bibfnamefont {J.~E.}\ \bibnamefont {Hoffman}},\ }\href {\doibase
  10.1063/1.3435466} {\bibfield  {journal} {\bibinfo  {journal} {Appl. Phys.
  Lett.}\ }\textbf {\bibinfo {volume} {96}},\ \bibinfo {pages} {213106}
  (\bibinfo {year} {2010})}\BibitemShut {NoStop}%
\bibitem [{\citenamefont {Oh}\ \emph {et~al.}(2010)\citenamefont {Oh},
  \citenamefont {Ko}, \citenamefont {Ramanathan},\ and\ \citenamefont
  {Cahill}}]{Oh2010}%
  \BibitemOpen
  \bibfield  {author} {\bibinfo {author} {\bibfnamefont {D.-W.}\ \bibnamefont
  {Oh}}, \bibinfo {author} {\bibfnamefont {C.}~\bibnamefont {Ko}}, \bibinfo
  {author} {\bibfnamefont {S.}~\bibnamefont {Ramanathan}}, \ and\ \bibinfo
  {author} {\bibfnamefont {D.~G.}\ \bibnamefont {Cahill}},\ }\href {\doibase
  10.1063/1.3394016} {\bibfield  {journal} {\bibinfo  {journal} {Appl. Phys.
  Lett.}\ }\textbf {\bibinfo {volume} {96}},\ \bibinfo {pages} {151906}
  (\bibinfo {year} {2010})}\BibitemShut {NoStop}%
\bibitem [{\citenamefont {Qazilbash}\ \emph {et~al.}(2009)\citenamefont
  {Qazilbash}, \citenamefont {Brehm}, \citenamefont {Andreev}, \citenamefont
  {Frenzel}, \citenamefont {Ho}, \citenamefont {Chae}, \citenamefont {Kim},
  \citenamefont {Yun}, \citenamefont {Kim}, \citenamefont {Balatsky},
  \citenamefont {Shpyrko}, \citenamefont {Maple}, \citenamefont {Keilmann},\
  and\ \citenamefont {Basov}}]{Qazilbash2009}%
  \BibitemOpen
  \bibfield  {author} {\bibinfo {author} {\bibfnamefont {M.}~\bibnamefont
  {Qazilbash}}, \bibinfo {author} {\bibfnamefont {M.}~\bibnamefont {Brehm}},
  \bibinfo {author} {\bibfnamefont {G.~O.}\ \bibnamefont {Andreev}}, \bibinfo
  {author} {\bibfnamefont {A.}~\bibnamefont {Frenzel}}, \bibinfo {author}
  {\bibfnamefont {P.-C.}\ \bibnamefont {Ho}}, \bibinfo {author} {\bibfnamefont
  {B.-G.}\ \bibnamefont {Chae}}, \bibinfo {author} {\bibfnamefont {B.-J.}\
  \bibnamefont {Kim}}, \bibinfo {author} {\bibfnamefont {S.}~\bibnamefont
  {Yun}}, \bibinfo {author} {\bibfnamefont {H.-T.}\ \bibnamefont {Kim}},
  \bibinfo {author} {\bibfnamefont {A.~V.}\ \bibnamefont {Balatsky}}, \bibinfo
  {author} {\bibfnamefont {O.}~\bibnamefont {Shpyrko}}, \bibinfo {author}
  {\bibfnamefont {M.~B.}\ \bibnamefont {Maple}}, \bibinfo {author}
  {\bibfnamefont {F.}~\bibnamefont {Keilmann}}, \ and\ \bibinfo {author}
  {\bibfnamefont {D.~N.}\ \bibnamefont {Basov}},\ }\href {\doibase
  10.1103/PhysRevB.79.075107} {\bibfield  {journal} {\bibinfo  {journal} {Phys.
  Rev. B}\ }\textbf {\bibinfo {volume} {79}},\ \bibinfo {pages} {075107}
  (\bibinfo {year} {2009})}\BibitemShut {NoStop}%
\bibitem [{\citenamefont {Leroux}\ \emph {et~al.}(1998)\citenamefont {Leroux},
  \citenamefont {Nihoul},\ and\ \citenamefont {{Van Tendeloo}}}]{Leroux1998}%
  \BibitemOpen
  \bibfield  {author} {\bibinfo {author} {\bibfnamefont {C.}~\bibnamefont
  {Leroux}}, \bibinfo {author} {\bibfnamefont {G.}~\bibnamefont {Nihoul}}, \
  and\ \bibinfo {author} {\bibfnamefont {G.}~\bibnamefont {{Van Tendeloo}}},\
  }\href {\doibase 10.1103/PhysRevB.57.5111} {\bibfield  {journal} {\bibinfo
  {journal} {Phys. Rev. B}\ }\textbf {\bibinfo {volume} {57}},\ \bibinfo
  {pages} {5111} (\bibinfo {year} {1998})}\BibitemShut {NoStop}%
\bibitem [{\citenamefont {Cao}\ \emph {et~al.}(2009)\citenamefont {Cao},
  \citenamefont {Ertekin}, \citenamefont {Srinivasan}, \citenamefont {Fan},
  \citenamefont {Huang}, \citenamefont {Zheng}, \citenamefont {Yim},
  \citenamefont {Khanal}, \citenamefont {Ogletree}, \citenamefont {Grossman},\
  and\ \citenamefont {Wu}}]{Cao2009e}%
  \BibitemOpen
  \bibfield  {author} {\bibinfo {author} {\bibfnamefont {J.}~\bibnamefont
  {Cao}}, \bibinfo {author} {\bibfnamefont {E.}~\bibnamefont {Ertekin}},
  \bibinfo {author} {\bibfnamefont {V.}~\bibnamefont {Srinivasan}}, \bibinfo
  {author} {\bibfnamefont {W.}~\bibnamefont {Fan}}, \bibinfo {author}
  {\bibfnamefont {S.}~\bibnamefont {Huang}}, \bibinfo {author} {\bibfnamefont
  {H.}~\bibnamefont {Zheng}}, \bibinfo {author} {\bibfnamefont {J.~W.~L.}\
  \bibnamefont {Yim}}, \bibinfo {author} {\bibfnamefont {D.~R.}\ \bibnamefont
  {Khanal}}, \bibinfo {author} {\bibfnamefont {D.~F.}\ \bibnamefont
  {Ogletree}}, \bibinfo {author} {\bibfnamefont {J.~C.}\ \bibnamefont
  {Grossman}}, \ and\ \bibinfo {author} {\bibfnamefont {J.}~\bibnamefont
  {Wu}},\ }\href {\doibase 10.1038/nnano.2009.266} {\bibfield  {journal}
  {\bibinfo  {journal} {Nat. Nanotechnol.}\ }\textbf {\bibinfo {volume} {4}},\
  \bibinfo {pages} {732} (\bibinfo {year} {2009})}\BibitemShut {NoStop}%
\bibitem [{\citenamefont {Takami}\ \emph
  {et~al.}(2012{\natexlab{b}})\citenamefont {Takami}, \citenamefont {Kanki},
  \citenamefont {Ueda}, \citenamefont {Kobayashi},\ and\ \citenamefont
  {Tanaka}}]{Takami2012}%
  \BibitemOpen
  \bibfield  {author} {\bibinfo {author} {\bibfnamefont {H.}~\bibnamefont
  {Takami}}, \bibinfo {author} {\bibfnamefont {T.}~\bibnamefont {Kanki}},
  \bibinfo {author} {\bibfnamefont {S.}~\bibnamefont {Ueda}}, \bibinfo {author}
  {\bibfnamefont {K.}~\bibnamefont {Kobayashi}}, \ and\ \bibinfo {author}
  {\bibfnamefont {H.}~\bibnamefont {Tanaka}},\ }\href {\doibase
  10.1103/PhysRevB.85.205111} {\bibfield  {journal} {\bibinfo  {journal} {Phys.
  Rev. B}\ }\textbf {\bibinfo {volume} {85}},\ \bibinfo {pages} {205111}
  (\bibinfo {year} {2012}{\natexlab{b}})}\BibitemShut {NoStop}%
\bibitem [{\citenamefont {Brassard}\ \emph {et~al.}(2005)\citenamefont
  {Brassard}, \citenamefont {Fourmaux}, \citenamefont {Jean-Jacques},
  \citenamefont {Kieffer},\ and\ \citenamefont {{El Khakani}}}]{Brassard2005}%
  \BibitemOpen
  \bibfield  {author} {\bibinfo {author} {\bibfnamefont {D.}~\bibnamefont
  {Brassard}}, \bibinfo {author} {\bibfnamefont {S.}~\bibnamefont {Fourmaux}},
  \bibinfo {author} {\bibfnamefont {M.}~\bibnamefont {Jean-Jacques}}, \bibinfo
  {author} {\bibfnamefont {J.~C.}\ \bibnamefont {Kieffer}}, \ and\ \bibinfo
  {author} {\bibfnamefont {M.~a.}\ \bibnamefont {{El Khakani}}},\ }\href
  {\doibase 10.1063/1.2001139} {\bibfield  {journal} {\bibinfo  {journal}
  {Appl. Phys. Lett.}\ }\textbf {\bibinfo {volume} {87}},\ \bibinfo {pages}
  {051910} (\bibinfo {year} {2005})}\BibitemShut {NoStop}%
\bibitem [{\citenamefont {Cao}\ \emph {et~al.}(2010{\natexlab{a}})\citenamefont
  {Cao}, \citenamefont {Gu}, \citenamefont {Fan}, \citenamefont {Chen},
  \citenamefont {Ogletree}, \citenamefont {Chen}, \citenamefont {Tamura},
  \citenamefont {Kunz}, \citenamefont {Barrett}, \citenamefont {Seidel},\ and\
  \citenamefont {Wu}}]{Cao2010}%
  \BibitemOpen
  \bibfield  {author} {\bibinfo {author} {\bibfnamefont {J.}~\bibnamefont
  {Cao}}, \bibinfo {author} {\bibfnamefont {Y.}~\bibnamefont {Gu}}, \bibinfo
  {author} {\bibfnamefont {W.}~\bibnamefont {Fan}}, \bibinfo {author}
  {\bibfnamefont {L.-Q.}\ \bibnamefont {Chen}}, \bibinfo {author}
  {\bibfnamefont {D.~F.}\ \bibnamefont {Ogletree}}, \bibinfo {author}
  {\bibfnamefont {K.}~\bibnamefont {Chen}}, \bibinfo {author} {\bibfnamefont
  {N.}~\bibnamefont {Tamura}}, \bibinfo {author} {\bibfnamefont
  {M.}~\bibnamefont {Kunz}}, \bibinfo {author} {\bibfnamefont {R.~C.}\
  \bibnamefont {Barrett}}, \bibinfo {author} {\bibfnamefont {J.}~\bibnamefont
  {Seidel}}, \ and\ \bibinfo {author} {\bibfnamefont {J.}~\bibnamefont {Wu}},\
  }\href {\doibase 10.1021/nl101457k} {\bibfield  {journal} {\bibinfo
  {journal} {Nano Lett.}\ }\textbf {\bibinfo {volume} {10}},\ \bibinfo {pages}
  {2667} (\bibinfo {year} {2010}{\natexlab{a}})}\BibitemShut {NoStop}%
\bibitem [{\citenamefont {Coy}\ \emph {et~al.}(2010)\citenamefont {Coy},
  \citenamefont {Cabrera}, \citenamefont {Sep{\'{u}}lveda},\ and\ \citenamefont
  {Fern{\`{a}}ndez}}]{Coy2010}%
  \BibitemOpen
  \bibfield  {author} {\bibinfo {author} {\bibfnamefont {H.}~\bibnamefont
  {Coy}}, \bibinfo {author} {\bibfnamefont {R.}~\bibnamefont {Cabrera}},
  \bibinfo {author} {\bibfnamefont {N.}~\bibnamefont {Sep{\'{u}}lveda}}, \ and\
  \bibinfo {author} {\bibfnamefont {F.}~\bibnamefont {Fern{\`{a}}ndez}},\
  }\href {\doibase 10.1063/1.3518508} {\bibfield  {journal} {\bibinfo
  {journal} {J. Appl. Phys.}\ }\textbf {\bibinfo {volume} {108}},\ \bibinfo
  {pages} {113115} (\bibinfo {year} {2010})}\BibitemShut {NoStop}%
\bibitem [{\citenamefont {Cao}\ \emph {et~al.}(2010{\natexlab{b}})\citenamefont
  {Cao}, \citenamefont {Fan}, \citenamefont {Zhou}, \citenamefont {Sheu},
  \citenamefont {Liu}, \citenamefont {Barrett},\ and\ \citenamefont
  {Wu}}]{Cao2010b}%
  \BibitemOpen
  \bibfield  {author} {\bibinfo {author} {\bibfnamefont {J.}~\bibnamefont
  {Cao}}, \bibinfo {author} {\bibfnamefont {W.}~\bibnamefont {Fan}}, \bibinfo
  {author} {\bibfnamefont {Q.}~\bibnamefont {Zhou}}, \bibinfo {author}
  {\bibfnamefont {E.}~\bibnamefont {Sheu}}, \bibinfo {author} {\bibfnamefont
  {A.}~\bibnamefont {Liu}}, \bibinfo {author} {\bibfnamefont {R.~C.}\
  \bibnamefont {Barrett}}, \ and\ \bibinfo {author} {\bibfnamefont
  {J.}~\bibnamefont {Wu}},\ }\href {\doibase 10.1063/1.3501052} {\bibfield
  {journal} {\bibinfo  {journal} {J. Appl. Phys.}\ }\textbf {\bibinfo {volume}
  {108}},\ \bibinfo {pages} {083538} (\bibinfo {year}
  {2010}{\natexlab{b}})}\BibitemShut {NoStop}%
\bibitem [{\citenamefont {Rúa}\ \emph {et~al.}(2012)\citenamefont {Rúa},
  \citenamefont {Cabrera}, \citenamefont {Coy}, \citenamefont {Merced},
  \citenamefont {Sepúlveda},\ and\ \citenamefont {Fernández}}]{Rua2012}%
  \BibitemOpen
  \bibfield  {author} {\bibinfo {author} {\bibfnamefont {A.}~\bibnamefont
  {Rúa}}, \bibinfo {author} {\bibfnamefont {R.}~\bibnamefont {Cabrera}},
  \bibinfo {author} {\bibfnamefont {H.}~\bibnamefont {Coy}}, \bibinfo {author}
  {\bibfnamefont {E.}~\bibnamefont {Merced}}, \bibinfo {author} {\bibfnamefont
  {N.}~\bibnamefont {Sepúlveda}}, \ and\ \bibinfo {author} {\bibfnamefont
  {F.~E.}\ \bibnamefont {Fernández}},\ }\href {\doibase 10.1063/1.4716191}
  {\bibfield  {journal} {\bibinfo  {journal} {J. Appl. Phys.}\ }\textbf
  {\bibinfo {volume} {111}},\ \bibinfo {pages} {104502} (\bibinfo {year}
  {2012})}\BibitemShut {NoStop}%
\bibitem [{\citenamefont {Rua}\ \emph {et~al.}(2010)\citenamefont {Rua},
  \citenamefont {Fern{\`{a}}ndez}, \citenamefont {Hines},\ and\ \citenamefont
  {Sep{\'{u}}lveda}}]{Rua2010}%
  \BibitemOpen
  \bibfield  {author} {\bibinfo {author} {\bibfnamefont {A.}~\bibnamefont
  {Rua}}, \bibinfo {author} {\bibfnamefont {F.}~\bibnamefont
  {Fern{\`{a}}ndez}}, \bibinfo {author} {\bibfnamefont {M.~A.}\ \bibnamefont
  {Hines}}, \ and\ \bibinfo {author} {\bibfnamefont {N.}~\bibnamefont
  {Sep{\'{u}}lveda}},\ }\href {\doibase 10.1063/1.3309749} {\bibfield
  {journal} {\bibinfo  {journal} {J. Appl. Phys.}\ }\textbf {\bibinfo {volume}
  {107}},\ \bibinfo {pages} {053528} (\bibinfo {year} {2010})}\BibitemShut
  {NoStop}%
\bibitem [{\citenamefont {Pellegrino}\ \emph {et~al.}(2012)\citenamefont
  {Pellegrino}, \citenamefont {Manca}, \citenamefont {Kanki}, \citenamefont
  {Tanaka}, \citenamefont {Biasotti}, \citenamefont {Bellingeri}, \citenamefont
  {Siri},\ and\ \citenamefont {Marr{\'{e}}}}]{Pellegrino2012}%
  \BibitemOpen
  \bibfield  {author} {\bibinfo {author} {\bibfnamefont {L.}~\bibnamefont
  {Pellegrino}}, \bibinfo {author} {\bibfnamefont {N.}~\bibnamefont {Manca}},
  \bibinfo {author} {\bibfnamefont {T.}~\bibnamefont {Kanki}}, \bibinfo
  {author} {\bibfnamefont {H.}~\bibnamefont {Tanaka}}, \bibinfo {author}
  {\bibfnamefont {M.}~\bibnamefont {Biasotti}}, \bibinfo {author}
  {\bibfnamefont {E.}~\bibnamefont {Bellingeri}}, \bibinfo {author}
  {\bibfnamefont {A.~S.}\ \bibnamefont {Siri}}, \ and\ \bibinfo {author}
  {\bibfnamefont {D.}~\bibnamefont {Marr{\'{e}}}},\ }\href {\doibase
  10.1002/adma.201104669} {\bibfield  {journal} {\bibinfo  {journal} {Adv.
  Mater.}\ }\textbf {\bibinfo {volume} {24}},\ \bibinfo {pages} {2929}
  (\bibinfo {year} {2012})}\BibitemShut {NoStop}%
\bibitem [{\citenamefont {Schlom}\ \emph {et~al.}(2008)\citenamefont {Schlom},
  \citenamefont {Chen}, \citenamefont {Pan}, \citenamefont {Schmehl},\ and\
  \citenamefont {Zurbuchen}}]{Schlom2008}%
  \BibitemOpen
  \bibfield  {author} {\bibinfo {author} {\bibfnamefont {D.~G.}\ \bibnamefont
  {Schlom}}, \bibinfo {author} {\bibfnamefont {L.-Q.}\ \bibnamefont {Chen}},
  \bibinfo {author} {\bibfnamefont {X.}~\bibnamefont {Pan}}, \bibinfo {author}
  {\bibfnamefont {A.}~\bibnamefont {Schmehl}}, \ and\ \bibinfo {author}
  {\bibfnamefont {M.~a.}\ \bibnamefont {Zurbuchen}},\ }\href {\doibase
  10.1111/j.1551-2916.2008.02556.x} {\bibfield  {journal} {\bibinfo  {journal}
  {J. Am. Ceram. Soc.}\ }\textbf {\bibinfo {volume} {91}},\ \bibinfo {pages}
  {2429} (\bibinfo {year} {2008})}\BibitemShut {NoStop}%
\bibitem [{\citenamefont {Okimura}\ and\ \citenamefont
  {Furumi}(2005)}]{Okimura2005}%
  \BibitemOpen
  \bibfield  {author} {\bibinfo {author} {\bibfnamefont {K.}~\bibnamefont
  {Okimura}}\ and\ \bibinfo {author} {\bibfnamefont {T.}~\bibnamefont
  {Furumi}},\ }\href {\doibase 10.1143/JJAP.44.3192} {\bibfield  {journal}
  {\bibinfo  {journal} {Jpn. J. Appl. Phys.}\ }\textbf {\bibinfo {volume}
  {44}},\ \bibinfo {pages} {3192} (\bibinfo {year} {2005})}\BibitemShut
  {NoStop}%
\bibitem [{\citenamefont {Muraoka}\ \emph {et~al.}(2002)\citenamefont
  {Muraoka}, \citenamefont {Ueda},\ and\ \citenamefont {Hiroi}}]{Muraoka2002d}%
  \BibitemOpen
  \bibfield  {author} {\bibinfo {author} {\bibfnamefont {Y.}~\bibnamefont
  {Muraoka}}, \bibinfo {author} {\bibfnamefont {Y.}~\bibnamefont {Ueda}}, \
  and\ \bibinfo {author} {\bibfnamefont {Z.}~\bibnamefont {Hiroi}},\ }\href
  {\doibase 10.1016/S0022-3697(02)00098-7} {\bibfield  {journal} {\bibinfo
  {journal} {J. Phys. Chem. Solids}\ }\textbf {\bibinfo {volume} {63}},\
  \bibinfo {pages} {965} (\bibinfo {year} {2002})}\BibitemShut {NoStop}%
\bibitem [{\citenamefont {Biasotti}\ \emph {et~al.}(2013)\citenamefont
  {Biasotti}, \citenamefont {Pellegrino}, \citenamefont {Buzio}, \citenamefont
  {Bellingeri}, \citenamefont {Bernini}, \citenamefont {Siri},\ and\
  \citenamefont {Marr{\'{e}}}}]{Biasotti2013}%
  \BibitemOpen
  \bibfield  {author} {\bibinfo {author} {\bibfnamefont {M.}~\bibnamefont
  {Biasotti}}, \bibinfo {author} {\bibfnamefont {L.}~\bibnamefont
  {Pellegrino}}, \bibinfo {author} {\bibfnamefont {R.}~\bibnamefont {Buzio}},
  \bibinfo {author} {\bibfnamefont {E.}~\bibnamefont {Bellingeri}}, \bibinfo
  {author} {\bibfnamefont {C.}~\bibnamefont {Bernini}}, \bibinfo {author}
  {\bibfnamefont {A.~S.}\ \bibnamefont {Siri}}, \ and\ \bibinfo {author}
  {\bibfnamefont {D.}~\bibnamefont {Marr{\'{e}}}},\ }\href {\doibase
  10.1088/0960-1317/23/3/035031} {\bibfield  {journal} {\bibinfo  {journal} {J.
  Micromechanics Microengineering}\ }\textbf {\bibinfo {volume} {23}},\
  \bibinfo {pages} {035031} (\bibinfo {year} {2013})}\BibitemShut {NoStop}%
\bibitem [{\citenamefont {Tselev}\ \emph {et~al.}(2011)\citenamefont {Tselev},
  \citenamefont {Budai}, \citenamefont {Strelcov}, \citenamefont {Tischler},
  \citenamefont {Kolmakov},\ and\ \citenamefont {Kalinin}}]{Tselev2011}%
  \BibitemOpen
  \bibfield  {author} {\bibinfo {author} {\bibfnamefont {A.}~\bibnamefont
  {Tselev}}, \bibinfo {author} {\bibfnamefont {J.~D.}\ \bibnamefont {Budai}},
  \bibinfo {author} {\bibfnamefont {E.}~\bibnamefont {Strelcov}}, \bibinfo
  {author} {\bibfnamefont {J.~Z.}\ \bibnamefont {Tischler}}, \bibinfo {author}
  {\bibfnamefont {A.}~\bibnamefont {Kolmakov}}, \ and\ \bibinfo {author}
  {\bibfnamefont {S.~V.}\ \bibnamefont {Kalinin}},\ }\href {\doibase
  10.1021/nl200493k} {\bibfield  {journal} {\bibinfo  {journal} {Nano Lett.}\
  }\textbf {\bibinfo {volume} {11}},\ \bibinfo {pages} {3065} (\bibinfo {year}
  {2011})}\BibitemShut {NoStop}%
\bibitem [{\citenamefont {Crunteanu}\ \emph {et~al.}(2010)\citenamefont
  {Crunteanu}, \citenamefont {Givernaud}, \citenamefont {Leroy}, \citenamefont
  {Mardivirin}, \citenamefont {Champeaux}, \citenamefont {Orlianges},
  \citenamefont {Catherinot},\ and\ \citenamefont {Blondy}}]{Crunteanu2010}%
  \BibitemOpen
  \bibfield  {author} {\bibinfo {author} {\bibfnamefont {A.}~\bibnamefont
  {Crunteanu}}, \bibinfo {author} {\bibfnamefont {J.}~\bibnamefont
  {Givernaud}}, \bibinfo {author} {\bibfnamefont {J.}~\bibnamefont {Leroy}},
  \bibinfo {author} {\bibfnamefont {D.}~\bibnamefont {Mardivirin}}, \bibinfo
  {author} {\bibfnamefont {C.}~\bibnamefont {Champeaux}}, \bibinfo {author}
  {\bibfnamefont {J.-C.}\ \bibnamefont {Orlianges}}, \bibinfo {author}
  {\bibfnamefont {A.}~\bibnamefont {Catherinot}}, \ and\ \bibinfo {author}
  {\bibfnamefont {P.}~\bibnamefont {Blondy}},\ }\href {\doibase
  10.1088/1468-6996/11/6/065002} {\bibfield  {journal} {\bibinfo  {journal}
  {Sci. Technol. Adv. Mater.}\ }\textbf {\bibinfo {volume} {11}},\ \bibinfo
  {pages} {065002} (\bibinfo {year} {2010})}\BibitemShut {NoStop}%
\bibitem [{\citenamefont {Ielmini}\ \emph {et~al.}(2007)\citenamefont
  {Ielmini}, \citenamefont {Lacaita},\ and\ \citenamefont
  {Mantegazza}}]{Ielmini2007}%
  \BibitemOpen
  \bibfield  {author} {\bibinfo {author} {\bibfnamefont {D.}~\bibnamefont
  {Ielmini}}, \bibinfo {author} {\bibfnamefont {A.~L.}\ \bibnamefont
  {Lacaita}}, \ and\ \bibinfo {author} {\bibfnamefont {D.}~\bibnamefont
  {Mantegazza}},\ }\href {\doibase 10.1109/TED.2006.888752} {\bibfield
  {journal} {\bibinfo  {journal} {IEEE Trans. Electron Devices}\ }\textbf
  {\bibinfo {volume} {54}},\ \bibinfo {pages} {308} (\bibinfo {year}
  {2007})}\BibitemShut {NoStop}%
\bibitem [{\citenamefont {Fan}\ \emph {et~al.}(2009)\citenamefont {Fan},
  \citenamefont {Huang}, \citenamefont {Cao}, \citenamefont {Ertekin},
  \citenamefont {Barrett}, \citenamefont {Khanal}, \citenamefont {Grossman},\
  and\ \citenamefont {Wu}}]{Fan2009}%
  \BibitemOpen
  \bibfield  {author} {\bibinfo {author} {\bibfnamefont {W.}~\bibnamefont
  {Fan}}, \bibinfo {author} {\bibfnamefont {S.}~\bibnamefont {Huang}}, \bibinfo
  {author} {\bibfnamefont {J.}~\bibnamefont {Cao}}, \bibinfo {author}
  {\bibfnamefont {E.}~\bibnamefont {Ertekin}}, \bibinfo {author} {\bibfnamefont
  {R.~C.}\ \bibnamefont {Barrett}}, \bibinfo {author} {\bibfnamefont {D.~R.}\
  \bibnamefont {Khanal}}, \bibinfo {author} {\bibfnamefont {J.~C.}\
  \bibnamefont {Grossman}}, \ and\ \bibinfo {author} {\bibfnamefont
  {J.}~\bibnamefont {Wu}},\ }\href {\doibase 10.1103/PhysRevB.80.241105}
  {\bibfield  {journal} {\bibinfo  {journal} {Phys. Rev. B}\ }\textbf {\bibinfo
  {volume} {80}},\ \bibinfo {pages} {241105} (\bibinfo {year}
  {2009})}\BibitemShut {NoStop}%
\bibitem [{\citenamefont {Rugar}\ \emph {et~al.}(2004)\citenamefont {Rugar},
  \citenamefont {Budakian}, \citenamefont {Mamin},\ and\ \citenamefont
  {Chui}}]{Rugar2004}%
  \BibitemOpen
  \bibfield  {author} {\bibinfo {author} {\bibfnamefont {D.}~\bibnamefont
  {Rugar}}, \bibinfo {author} {\bibfnamefont {R.}~\bibnamefont {Budakian}},
  \bibinfo {author} {\bibfnamefont {H.~J.}\ \bibnamefont {Mamin}}, \ and\
  \bibinfo {author} {\bibfnamefont {B.~W.}\ \bibnamefont {Chui}},\ }\href
  {\doibase 10.1038/nature02658} {\bibfield  {journal} {\bibinfo  {journal}
  {Nature}\ }\textbf {\bibinfo {volume} {430}},\ \bibinfo {pages} {329}
  (\bibinfo {year} {2004})}\BibitemShut {NoStop}%
\bibitem [{\citenamefont {O'Connell}\ \emph {et~al.}(2010)\citenamefont
  {O'Connell}, \citenamefont {Hofheinz}, \citenamefont {Ansmann}, \citenamefont
  {Bialczak}, \citenamefont {Lenander}, \citenamefont {Lucero}, \citenamefont
  {Neeley}, \citenamefont {Sank}, \citenamefont {Wang}, \citenamefont {Weides},
  \citenamefont {Wenner}, \citenamefont {Martinis},\ and\ \citenamefont
  {Cleland}}]{OConnell2010}%
  \BibitemOpen
  \bibfield  {author} {\bibinfo {author} {\bibfnamefont {A.~D.}\ \bibnamefont
  {O'Connell}}, \bibinfo {author} {\bibfnamefont {M.}~\bibnamefont {Hofheinz}},
  \bibinfo {author} {\bibfnamefont {M.}~\bibnamefont {Ansmann}}, \bibinfo
  {author} {\bibfnamefont {R.~C.}\ \bibnamefont {Bialczak}}, \bibinfo {author}
  {\bibfnamefont {M.}~\bibnamefont {Lenander}}, \bibinfo {author}
  {\bibfnamefont {E.}~\bibnamefont {Lucero}}, \bibinfo {author} {\bibfnamefont
  {M.}~\bibnamefont {Neeley}}, \bibinfo {author} {\bibfnamefont
  {D.}~\bibnamefont {Sank}}, \bibinfo {author} {\bibfnamefont {H.}~\bibnamefont
  {Wang}}, \bibinfo {author} {\bibfnamefont {M.}~\bibnamefont {Weides}},
  \bibinfo {author} {\bibfnamefont {J.}~\bibnamefont {Wenner}}, \bibinfo
  {author} {\bibfnamefont {J.~M.}\ \bibnamefont {Martinis}}, \ and\ \bibinfo
  {author} {\bibfnamefont {A.~N.}\ \bibnamefont {Cleland}},\ }\href {\doibase
  10.1038/nature08967} {\bibfield  {journal} {\bibinfo  {journal} {Nature}\
  }\textbf {\bibinfo {volume} {464}},\ \bibinfo {pages} {697} (\bibinfo {year}
  {2010})},\ \Eprint {http://arxiv.org/abs/1602.03841} {arXiv:1602.03841}
  \BibitemShut {NoStop}%
\bibitem [{\citenamefont {Naik}\ \emph {et~al.}(2009)\citenamefont {Naik},
  \citenamefont {Hanay}, \citenamefont {Hiebert}, \citenamefont {Feng},\ and\
  \citenamefont {Roukes}}]{Naik2009a}%
  \BibitemOpen
  \bibfield  {author} {\bibinfo {author} {\bibfnamefont {A.~K.}\ \bibnamefont
  {Naik}}, \bibinfo {author} {\bibfnamefont {M.~S.}\ \bibnamefont {Hanay}},
  \bibinfo {author} {\bibfnamefont {W.~K.}\ \bibnamefont {Hiebert}}, \bibinfo
  {author} {\bibfnamefont {X.~L.}\ \bibnamefont {Feng}}, \ and\ \bibinfo
  {author} {\bibfnamefont {M.~L.}\ \bibnamefont {Roukes}},\ }\href {\doibase
  10.1038/nnano.2009.152} {\bibfield  {journal} {\bibinfo  {journal} {Nat.
  Nanotechnol.}\ }\textbf {\bibinfo {volume} {4}},\ \bibinfo {pages} {445}
  (\bibinfo {year} {2009})}\BibitemShut {NoStop}%
\bibitem [{\citenamefont {Sengupta}\ \emph {et~al.}(2010)\citenamefont
  {Sengupta}, \citenamefont {Solanki}, \citenamefont {Singh}, \citenamefont
  {Dhara},\ and\ \citenamefont {Deshmukh}}]{Sengupta2010}%
  \BibitemOpen
  \bibfield  {author} {\bibinfo {author} {\bibfnamefont {S.}~\bibnamefont
  {Sengupta}}, \bibinfo {author} {\bibfnamefont {H.~S.}\ \bibnamefont
  {Solanki}}, \bibinfo {author} {\bibfnamefont {V.}~\bibnamefont {Singh}},
  \bibinfo {author} {\bibfnamefont {S.}~\bibnamefont {Dhara}}, \ and\ \bibinfo
  {author} {\bibfnamefont {M.~M.}\ \bibnamefont {Deshmukh}},\ }\href {\doibase
  10.1103/PhysRevB.82.155432} {\bibfield  {journal} {\bibinfo  {journal} {Phys.
  Rev. B}\ }\textbf {\bibinfo {volume} {82}},\ \bibinfo {pages} {155432}
  (\bibinfo {year} {2010})}\BibitemShut {NoStop}%
\end{thebibliography}%

\end{document}